\documentclass[aps,prd,floatfix,twocolumn,letterpaper,superscriptaddress]{revtex4}

\usepackage{bm, graphicx, color}
\usepackage{amsmath,amssymb,amsfonts}
\usepackage{ wasysym }

\newcommand{\HH}{\mathcal{H}}

\begin{document}
\title{Gravitational mass of relativistic matter and antimatter}

\author{Tigran Kalaydzhyan}
\email[]{tigran@uic.edu}
\affiliation{Department of Physics, University of Illinois, Chicago, Illinois 60607-7059, USA}
\affiliation{Department of Physics and Astronomy, Stony Brook University,\\ Stony Brook, New York 11794-3800, USA}

\begin{abstract}
The universality of free fall, the weak equivalence principle (WEP), is a cornerstone of the general theory of relativity, the most precise theory of gravity confirmed in all experiments up to date. The WEP states the equivalence of the inertial, $m$, and gravitational, $m_g$, masses and was tested in numerous occasions with normal matter
at relatively low energies. However, there is no confirmation for the matter and antimatter at high energies.
For the antimatter the situation is even less clear -- current direct observations of trapped antihydrogen suggest the limits $-65 < m_g / m < 110$ not excluding the so-called antigravity phenomenon, i.e. repulsion of the antimatter by Earth. Here we demonstrate an indirect bound $0.96 < m_g/m < 1.04$
on the gravitational mass of relativistic electrons and positrons
coming from the absence of the vacuum Cherenkov radiation at the Large Electron-Positron Collider (LEP) and stability of photons at the Tevatron collider in presence of the annual variations of the solar gravitational potential. Our result clearly rules out the speculated antigravity.
By considering the absolute potential of the Local Supercluster (LS), we also predict the bounds $1 - 4\times 10^{-7} < m_g/m < 1 + 2\times 10^{-7}$ for an electron and positron.
 Finally, we comment on a possibility of performing complementary tests at the future International Linear Collider (ILC) and Compact Linear Collider (CLIC).
\end{abstract}
\maketitle

{\it Introduction.---} Since the formulation of the general relativity (GR) by Einstein in 1915-16~\cite{Einstein:1915ca, Einstein:1916vd}
 there were numerous tests confirming validity of the theory with an exceptional precision~\cite{Will:2014xja}. The weak equivalence principle (WEP), postulating the universality of the free fall, or equivalence of the inertial and gravitational masses, was confirmed in torsion balance experiments~\cite{Wagner:2012ui} at the $2\times 10^{-13}$ level for the normal matter. The idea of ``antigravity'' for an exotic matter seems to exist since the end of the XIX century~\cite{Hicks}, where it appeared together with the idea of antimatter. The modern, quantum, concept of antimatter begins with the theoretical paper of Dirac~\cite{Dirac:1928hu} in 1928 and experimental observation of antielectron (positron) by Anderson~\cite{Anderson:1933mb} in 1933. However, since then, there is no conclusion made about the gravitational interaction of antimatter~\cite{Fischler:2008zz}. The most precise direct observation of cold-trapped antihydrogen~\cite{Amole:2013gma} sets the limits on the ratio between the inertial $m$ and gravitational $m_g$ masses of the antihydrogen, $-65 < m_g / m < 110$, including systematic errors, at the 5\% significance level~\cite{Amole:2013gma}. At the same time, indirect limits have a long history and are much stricter (even though, most of them use additional assumptions), see Review~\cite{Nieto:1991xq} for the arguments prior to 1991. At the moment, the most precise bounds on the difference between the gravitational masses of the matter and antimatter (to our knowledge) are obtained from the comparison of decay parameters of the $K^0 - \bar{K}^0$ system~\cite{Apostolakis:1999tk} ($1.8\times 10^{-9}$ level with gravitational potential variations and $1.9\times 10^{-14}$ with the LS potential) and from comparison of cyclotron frequencies~\cite{Gabrielse:1999kc} of the $p - \bar p$ system~\cite{Hughes:1990ay} ($10^{-6}$ level with LS potential). Equality of the inertial masses for the considered (anti)particles is supported by the $\mathcal{CPT}$-symmetry tested with a much higher precision~\cite{Agashe:2014kda}.
 These and other indirect limits are, however, not absolute, but relative (between particles and antiparticles) and for relatively low energies. There is, therefore, no guarantee that, e.g., the strange matter (kaons) at any energies, or normal matter and antimatter at high energies (several GeV and higher) will obey WEP. These limits also do not restrict certain WEP violation models, such as the ``isotropic parachute model''~\cite{Tasson:2015cqa}.

 Even though astrophysical tests of the Lorentz invariance~\cite{Liberati:2013xla, Mattingly:2005re,  Tasson:2014dfa, Bluhm:2005uj} can be, perhaps,
  used for the precise tests of the WEP (mainly for electrons and protons), they rely on certain models describing the high-energy sources and their dynamics. It is, therefore, desirable to obtain similar or better constrains in a well-controlled experimental setup.

In this paper, we constrain possible deviations from WEP for ultrarelativistic electrons and positrons
based on the absence of the vacuum Cherenkov radiation from 104.5~GeV electrons and positrons at the LEP at CERN, and on the absence of the photon decays for 340.5~GeV photons at the Tevatron accelerator at Fermilab. It is known that the large Lorentz $\gamma$-factor for the ultrarelativistic particles reveals certain gravity and Lorentz-violating effects~\cite{Altschul:2009xh,Gharibyan:2012gp, Kalaydzhyan}, and suppresses the ordinary electromagnetic interaction~\cite{Chao:1993zn} otherwise overwhelming gravitational forces~\cite{fairbank}. This nontrivial fact makes accelerator experiments suitable for the gravitational studies. In addition, continuous collection of the accelerator data makes it possible to study changes in the observables (or exclusion regions in the parameter space) relative to the periodic variations of the astrophysical potentials. This gives one an opportunity to avoid assumptions on the absolute values of the gravitational potentials~\cite{Apostolakis:1999tk}.
An additional advantage of the vacuum Cherenkov radiation for the positron (electron) is its independence of the gravitational properties of the electron (positron). We also choose the electron and positron for our studies because of the absence of an additional internal structure or flavor composition, avoiding possible speculations on, e.g., undiscovered ``strange'', ``isotopic'' or ``hypercharge'' forces~\cite{Bernstein:1964hh, Bell:1964ff}.

{\it Dispersion relations.---} Let us begin with a description of the gravity effects on the high-energy processes.
Gravitational field of the Earth (Sun or other distant massive celestial objects) around the accelerator can be considered homogeneous and described by an isotropic metric for a static weak field, 
\begin{align}
ds^2 = \HH^2 dt^2 - \HH^{-2}(dx^2 + dy^2 + dz^2)\,,\label{metric}
\end{align}
where $\HH^2 = 1+2\Phi$, and $\Phi$ is the gravitational potential, defining the acceleration of free-falling bodies, $\textbf{a} = - \mathbf{\nabla} \Phi (\textbf{x})$, taken at the Earth's surface~\footnote{The formula for acceleration holds in the nonrelativistic case as well as in the relativistic case if the gravitational forces act perpendicular to the velocity of the particle. If the gravitational force $\textbf{F} = - \gamma m_g \mathbf{\nabla} \Phi$ was parallel to the velocity of the particle $\textbf{v}_m$, then it would contribute to the acceleration $\textbf{a}$ with an additional factor $1/\gamma^2$, i.e.,
$\textbf{a}=(\textbf{F}-(\textbf{v}_m\cdot \textbf{F})\textbf{v}_m)/(\gamma m)$, see Ref.~\cite{Landau1975}.
}.
Here and after we work in natural units, $c=\hbar=1$. We assume that the metric (\ref{metric}) results from a nonrelativistic distribution of normal matter (which is true in the cases considered below), for which the WEP holds with a high precision. Therefore, there is no difference between the inertial and gravitational masses appearing in Eq.~(\ref{metric}).

For a massive probe relativistic particle or antiparticle of inertial mass $m$ and gravitational mass ${m_g}$ (assuming one does not know if they are equal \textit{a priori}), we can write the gravitational potential as
\begin{align}
\Phi_m = \Phi \,\frac{{m_g}}{m}\,, \qquad \HH_m^2 \equiv 1+ 2\Phi_m\,.
\end{align}
This gravitational potential does not appear as a solution to Einstein's equations, but is a way of generalizing the gravitational coupling of the probe massive particles to the background which reproduces the Newton's gravitational law and its relativistic extension~\cite{Landau1975}. Particles participating in high-energy experiments considered below can be treated as probe particles due to their negligible masses and energies, comparing to the ones of the astrophysical objects creating the background (\ref{metric}). We also do not have a goal of suggesting an alternative action-based theory of gravity, e.g., to take into account the backreaction of the antimatter, since this is not needed with the assumptions made in the paper.

Let us consider a photon with coordinate 4-momentum $\tilde k_\mu = (\tilde\omega, \tilde{\textbf{k}})$, and a massive ultrarelativistic particle with coordinate 4-momentum $\tilde p_\mu = (\tilde {\cal E}, \tilde{\textbf{p}})$ and mass $m  \ll \tilde {\cal E}$. The metric (\ref{metric}) modifies the coordinate speed of light,
\begin{align}
v_\gamma \equiv |d \textbf{x} / d t| = \HH^2\,,
\end{align}
which can be obtained from the null geodesics, $d s^2=0$, defining the photon's trajectory. For a massive probe particle moving with the coordinate speed $\tilde{\textbf{v}}_m$, the line element can be rewritten then as
\begin{align}
ds^2 = \HH_m^2 \left( 1- \HH_m^{-4} \tilde{\textbf{v}}_m^2 \right) dt^2,
\end{align}
and the relativistic action takes the form
\begin{align}
S = -\int m\,ds = -\int m \HH_m \sqrt{1-\HH_m^{-4} \tilde{\textbf{v}}_m^2} dt\,.
\end{align}
Using this action, one can easily obtain the coordinate momentum $\tilde{\textbf{p}}$ and the Hamiltonian (energy) $\tilde{{\cal E}}$,
\begin{align}
 \tilde{\textbf{p}} = \frac{m \HH_m^{-3}}{\sqrt{1-\HH_m^{-4} \tilde{\textbf{v}}_m^2}}\, \tilde{\textbf{v}}_m, \qquad \tilde{{\cal E}} = \frac{m \HH_m}{\sqrt{1-\HH_m^{-4} \tilde{\textbf{v}}_m^2}}\,.\label{energy}
\end{align}
The modified coordinate dispersion relations for the photon and a massive particle is given then by
\begin{align}
\tilde k^2 = \HH^{-4} \tilde \omega^2,\quad\tilde p^2 = \left(1+ 4 |\Phi| \frac{{m_g}}{m} \right) \left(\tilde{\cal E}^2- m^2\right)\,,
\end{align}
where $\tilde k = |\tilde{\textbf{k}}|$, $\tilde p = |\tilde{\textbf{p}}|$ and we use $|\Phi|$ instead of $-\Phi$ for the convenience (since potentials of massive bodies are usually taken negative in a coordinate system with the origin in the center of these bodies).
The physical expressions can be obtained from the coordinate ones by rescaling, $\textbf{v} = \HH^{-2} \tilde{\textbf{v}}$, $k = \HH \tilde k$, $\omega = \HH^{-1} \tilde\omega$, $p = \HH \tilde p$, ${\cal E} = \HH^{-1} \tilde{{\cal E}}$, and absorbing the $\HH$ factors in (\ref{metric}) into the definitions of the coordinates. We also assume that there is no modification of the physical speed of light within the considered accuracy~\cite{Will:2014xja,Liberati:2013xla,AmelinoCamelia:2008qg}. Finally, the physical momenta of the photon and the massive particle take the form 
\begin{align}
k = \omega, \qquad ~~ p = {\cal E} \left(1+ 2 |\Phi| \frac{\Delta m}{m} \right) \sqrt{1- \frac{m^2}{{\cal E}^2}}\,,\label{dispersion}
\end{align}
where $\Delta m = {m_g} - m$, and we treat $\kappa \equiv 2 |\Phi| \Delta m / m$ as a small parameter. Physically, the obtained expressions demonstrate an anomalous redshift the massive particle would get if WEP was violated.
This form of the dispersion relations is similar to the ones used in the phenomenology and tests of the quantum gravity and Lorentz violation~\cite{AmelinoCamelia:2008qg, Hohensee:2008xz, Hohensee:2009zk, Klinkhamer:2008ky, Altschul:2009xh}. For instance, the dispersion relations (\ref{dispersion}) can be obtained from
the minimal Lorentz-violating Standard Model Extension (SME)~\cite{sme} with parameters $c_{00}=3c_{ii}=3\kappa /4$ (no summation by $i$) and other parameters set to zero. With the assumption of universality of the speed of light, this is a reasonable approximation as soon as $|\kappa| > 10^{-13}$, which corresponds to the upper boundary on the next dominating SME parameter~\cite{Kostelecky:2008ts}.
 Therefore, one can use known tests of the Lorentz-violation (e.g., vacuum Cherenkov radiation, photon decay, synchrotron losses and others) to obtain limits on the parameter $\kappa$ and, hence, the difference between the gravitational and inertial masses. One of such tests is presented in details in Refs.~\cite{Hohensee:2009zk, Hohensee:2008xz} (our $\kappa$ can be treated as equivalent to their $4c_{00}/3 - \tilde \kappa_{\mathrm{tr}}$).

\begin{figure}
\centering
\includegraphics[width=3.5cm]{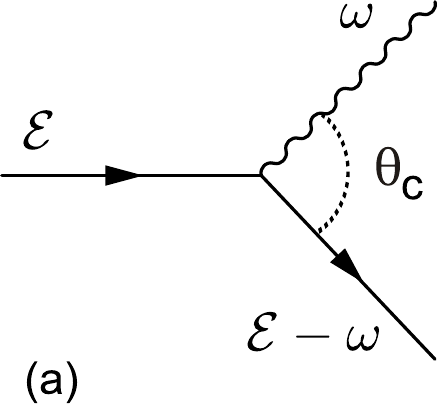}\hspace{1cm}
\includegraphics[width=3.5cm]{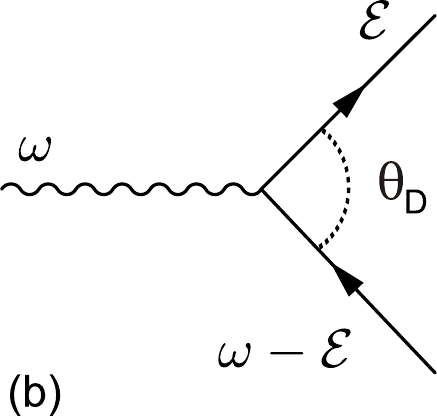}
\caption{Left: vacuum Cherenkov radiation, $e^\pm \rightarrow e^\pm \gamma$ (electron or positron of energy ${\cal E}$ emits a photon of energy $\omega$ with an angle between products $\theta_C$). Right: photon decay, $\gamma \rightarrow e^+ e^-$ (photon of energy $\omega$ emits an electron of energy ${\cal E}$ and positron of energy $\omega - {\cal E}$ with an angle between them $\theta_D$).\label{diagrams}}
\end{figure}

{\it Vacuum Cherenkov radiation. ---} On-shell emission of a photon by an electron or positron in the vacuum, so-called vacuum Cherenkov radiation, is normally forbidden kinematically. 
However, in the presence of the nontrivial modification of the dispersion relation (\ref{dispersion}) with $\kappa < 0$, the energy-momentum conservation condition allows such a process in a certain range of the angles $\theta_C$, see Fig.~\ref{diagrams}(a). In other words, the electron (positron) is allowed to move faster than light at a certain energy. The energy threshold ${\cal E}_\mathrm{th}$ and the emission rate $\Gamma_C$ are given then by~\cite{Altschul:2007tn}
\begin{align}
{\cal E}_\mathrm{th} = \frac{m_e}{\sqrt{- 2 \kappa}},\qquad \Gamma_C = \alpha\, m_e^2\, \frac{({\cal E} - {\cal E}_\mathrm{th})^2}{2 {\cal E}^3}\,,
\end{align}
where $m_e$ is the inertial electron (positron) mass and $\alpha$ is the fine-structure constant.
Due to the high emission rate,
 a particle above ${\cal E}_\mathrm{th}$ will be rapidly slowed down to the threshold energy through the photon radiation. For instance, the positrons at LEP at CERN, with the energies ${\cal E}=104.5$ GeV and the arbitrarily chosen threshold energy ${\cal E}_\mathrm{th}=100$ GeV would be decelerated to the subluminal speeds just within 1.2~cm of travel~\cite{Hohensee:2009zk} (compare to, e.g., $\sim 6$~km distance between LEP accelerating RF systems~\cite{LEP}).
 Since this was never observed, ${\cal E}_\mathrm{th}> 100$ GeV and $\kappa > \kappa^- = -1.31\times 10^{-11}$. A more rigorous analysis done in Ref.~\cite{Hohensee:2008xz} shows that the energy loss due to the Cherenkov radiation at given threshold would be much larger than the one actually allowed by the measurements (the relative error on the energy determination for the majority of LEP 2 running is $1.2\times 10^{-4}$~\cite{Assmann:2004gc}). This method has an advantage of not using a comparison between properties of electron and positron (and thus exploiting $\mathcal{CPT}$-invariance) but limiting the gravitational mass of the electron and positron directly.

{\it Vacuum photon decay. ---} As another standard textbook example, decay of a photon into an electron-positron pair [Fig.~\ref{diagrams}(b)] is also forbidden kinematically, since $\cos \theta_D > 1$ can be never satisfied. However, at $\kappa > 0$ it becomes possible. The threshold on the photon energy $\omega_\mathrm{th}$ and the decay rate $\Gamma_D$ are given by~\cite{Coleman:1998ti, Klinkhamer:2008ky, Hohensee:2009zk}
\begin{align}
&\omega_\mathrm{th} = \sqrt{\frac{2}{\kappa}} m_e, \label{rate}\\
 &\Gamma_D = \frac{2}{3}\alpha\, \omega\, \frac{m_e^2}{\omega_\mathrm{th}^2}\left(2+\frac{\omega_\mathrm{th}^2}{\omega^2} \right)\sqrt{1-\frac{\omega_\mathrm{th}^2}{\omega^2}}\,,\label{gdecay}
\end{align}
where we assumed for simplicity that the electron's dispersion relation is modified in the same way as positron's, since there are no precise limits on the gravitational mass of the ultrarelativistic electron either. If the electron is assumed to obey WEP and hold the standard dispersion relation, then the values (\ref{rate}, \ref{gdecay}) will be slightly modified (e.g., $\omega_\mathrm{th}$ will change by a factor $\sqrt{2}$).
Following Refs.~\cite{Hohensee:2009zk, Hohensee:2008xz}, we consider isolated photon production with an associated jet, $p\bar p \rightarrow \gamma + \mathrm{jet} + X$, as measured by the D0 detector~\cite{Abazov:2008er} at Fermilab Tevatron collider at the center-of-mass energy $\sqrt{s}=1.96$~TeV. The photons up to energies 340.5~GeV were observed~\cite{Abachi:1993em} and we conservatively take the lower bound, 300~GeV, of the considered 340.5~GeV bin.
The possible photon decay process is very efficient and leads to a fast energy loss. As an example, 300~GeV photons with an energy 1\% above threshold would decay after traveling an average distance of only 0.1~mm (for comparison, the photons should travel a minimal distance of 78~cm in order to be measured by the central calorimeter of the D0 detector~\cite{Abachi:1993em}). As shown in Ref.~\cite{Hohensee:2008xz}, the hypothetical photon decay at 300~GeV would lead to the deficit in the photon flux much larger than the one allowed by the difference between the QCD predictions and experimental data~\cite{Abachi:1993em}.
This leads to the right bound $\kappa < \kappa^+ = 5.80\times 10^{-12}$. Possible modification of (\ref{rate}) discussed above could be considered as making the bound less precise. However, isolated photons with energies up to 1~TeV were observed in $\sqrt{s}=7$ TeV $pp$-collisions at the Large Hadron Collider~\cite{ATLAS:2013ema} (LHC) at CERN. The photon flux there is well described by theoretical predictions~\cite{ATLAS:2013ema}, making our bound even more conservative.

{\it Results. ---}
Using the thresholds from the previous sections, we impose the limits on the deviation $\Delta m_e$ of the positron's (electron's) gravitational mass $m_{e,g}$ from the inertial mass $m_e$,
\begin{align}
-\frac{m_e^2}{4 {\cal E}_\mathrm{th}^2 |\Phi|} < \frac{\Delta m_{e}}{m_e} < \frac{m_e^2}{\omega_\mathrm{th}^2 |\Phi|}\,.\label{limits}
\end{align}
As a consequence of the deviation from the equivalence principle, the absolute values of the gravitational potentials start playing a role. The total potential can be written as
\begin{align}
\Phi = \Phi_\oplus + \Phi_{\leftmoon} + \Phi_\odot + \Phi_{\mathrm{MW}} + \Phi_{\mathrm{SC}} + \Phi_{\mathrm{U}} + C,\label{potent}
\end{align}
i.e. a sum of the gravitational potentials of the Earth, Moon, Sun, Milky Way, rest of the Local Supercluster, rest of the Universe and a constant $C$ (assuming it being small, so the Newtonian limit can be applied), respectively. The largest known contribution at the surface of the Earth is the potential of the Local Supercluster with $|\Phi_\mathrm{SC}| \simeq 3\times 10^{-5}$ (compare to the Earth's $|\Phi_\oplus| = G M_\oplus / R_\oplus = 7 \times 10^{-10}$ and Sun's $|\Phi_\odot| = 9.9\times 10^{-10}$). Taking this value of the potential, we obtain the numerical limits on the gravitational mass,
\begin{align}
1 - 4\times 10^{-7} < m_{e, g}/m_e < 1 + 2\times 10^{-7}\,, \label{numlimits}
\end{align}
supporting the WEP for the antimatter.  Taken that the current estimates on the minimal range of the gravitational forces is about $100$~Mpc, see Ref.~\cite{Choudhury:2002pu}, one can improve our estimates by taking into account gravitational potentials from larger or more distant mass distributions.

The potential problem is the values of $\Phi_{\mathrm{U}}$ and $C$. If all the matter in the Universe contributes to the total potential in the same way as the observable matter, it can increase the value of $|\Phi|$ used for the estimates in (\ref{limits}) and make our bounds stronger. However, the value of $C$ depends on the current or future cosmological model and is not known \textit{a priori}. If it contributes with an opposite sign
and reduces the given potential value by one or several order of magnitude, then our estimates may not be correct. This would, however, introduce a fine-tuning, meaning our Galaxy and surrounding neighborhood have a privileged position in the Universe, such that the constant $C$ defined by large scale structures in the Universe cancels out the effect of the Local Supercluster ~\cite{Good:1961zz}. Finally, if $C$ changes the sign of the given potential without reducing the absolute value, it will only change the orders of the two-sided bound (\ref{limits}).

In order to avoid the problem of using the absolute potentials (\ref{potent}), one can consider periodic (daily, monthly, annual etc.) variations of the astrophysical potentials while the experiments are performed. Taking the two-sided bound $\kappa^- < \kappa < \kappa^+$ for two potentials, $\Phi$ and $\Phi + \Delta\Phi$ (e.g. both the vacuum Cherenkov radiation and photon decay were absent during the experiment), one can easily deduce
\begin{align}
\kappa^- - \kappa^+ < 2 \Delta\Phi\, \frac{\Delta m_{e}}{m_e} < \kappa^+ - \kappa^-\,. \label{newlimits}
\end{align}
Leading contribution to the variation of the total potential (\ref{potent}) within a few months time is given by
\begin{align}
\Delta\Phi = - \Phi_\odot \frac{\Delta d_{SE}}{d_{SE}}\,,
\end{align}
where $\Delta d_{SE}$ is the variation of the distance between Sun and Earth, $d_{SE}$, due to the eccentricity of the Earth's orbit. Considering the time interval of the LEP 104.5~GeV operation in 2000 from the beginning of April until the shutdown on November 2nd~\cite{Assmann:2001bh}, one can estimate the maximal variation $\Delta d_{SE} \approx 2.46\times 10^{-2} \mathrm{AU}$, which can be obtained from, e.g., the NASA's Jet Propulsion Laboratory (JPL) solar system data~\cite{jpl}. This gives the maximal variation of the potential, $|\Delta \Phi| = 2.43\times 10^{-10}$. Data from the D0 detector at Tevatron~\cite{Abazov:2008er} used for the photon decay analysis~\cite{Hohensee:2008xz} was collected for several years~\cite{Andeen:2007zc} covering the Earth-Sun distance changes related to the LEP data. Therefore, using (\ref{newlimits}) and the value of $|\Delta \Phi|$, we obtain
\begin{align}
\left|\frac{\Delta m_{e}}{m_e}\right| < 0.0389\,,\label{relmlimit}
\end{align}
i.e. a 4\% limit on the possible deviation. One may argue that the binning of the data (e.g. month-to-month) is required. However, taken that considered effects are so drastic, they would be readily visible on the initial stages of the data analysis. Total error, coming mainly from the precision at which the gravitational potentials are taken, gives up to 2\% uncertainty to (\ref{relmlimit}) and up to 35\% uncertainty to (\ref{numlimits}), which is reflected in the number of presented significant digits.

\begin{figure}[t!]
\centering
\includegraphics[width=7cm]{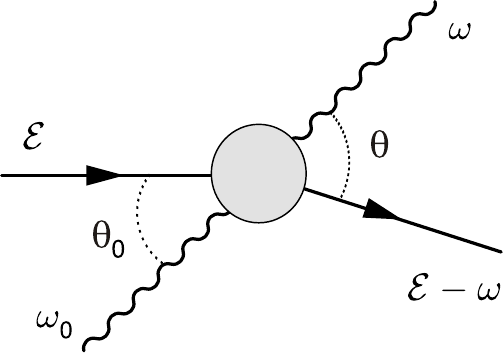}
\caption{Kinematics of the Compton scattering (a photon of energy $\omega_0$ scatters off an electron or positron of energy ${\cal E}$ and acquires energy $\omega$).\label{compton}}
\end{figure}

{\it Additional future prospects. ---} Complementary limits of a similar or higher precision can be obtained from the shift of the edge of the Compton spectrum in high-energy laser Compton scattering~\cite{Kalaydzhyan}. The process is depicted in Fig.~\ref{compton} and consists of a photon of energy $\omega_0$ colliding with an electron (or positron) of energy ${\cal E} \gg m_e$ under angle $\theta_0$ and scattering off under angle $\theta$ with respect to the resulting electron (positron). If the acquired energy of the photon $\omega$ is close to the maximal possible $\omega_{max}$ (the Compton edge), then the scattering angle is small, $\theta \ll 1$. If further $\omega_0 \ll {\cal E}$, then the energy-momentum conservation condition with inserted modified dispersion relations (\ref{dispersion}) leads to an expression~\cite{Gharibyan:2012gp, Kalaydzhyan}
\begin{align}
 \kappa = \frac{m_e^2}{2 {\cal E} (\omega-{\cal E})}\left( 1+x+\left(\frac{{\cal E}-\omega}{m_e} \right)^2\theta^2 - x \frac{{\cal E}}{\omega}\right),\label{index}
\end{align}
where $x \equiv 4{\cal E} \omega_0\sin^2{(\theta_0/2)}/m_e^2$ is a kinematic parameter defined by the experimental setup. If $\kappa=0$, i.e. there is no violation of WEP, then the nominal value of the Compton edge (at $\theta=0$) is
\begin{align}
\omega_{max}^{(\kappa=0)} = \frac{\mathcal{E}x}{1+x}\,. \label{nom}
\end{align}
If, however, there is a small deviation from the equivalence principle, the Compton edge will be shifted by a value $\Delta\omega \ll \omega_{max}$,
\begin{align}
\omega_{max} = \frac{\mathcal{E}x}{1+x}+ \Delta\omega\,.
\end{align}
Substituting the definition of $\Delta\omega$ into (\ref{index}) at $\theta=0$, we obtain
\begin{align}
 \frac{\Delta\omega}{\omega_{max}} = \frac{4 {\cal E}^2 |\Phi|}{m_e^2 (1+x)^2}\cdot \frac{\Delta m_{e}}{m_e}\,,\label{wtom}
\end{align}
To demonstrate the sensitivity of the process, let us consider a high energy $e^-/e^+$ beam with ${\cal E}=250$ GeV planned to be initially generated at ILC~\cite{Behnke:2013lya}. The Compton scattering facilities at ILC will be operated for polarimetry~\cite{Behnke:2013lya} with a typical setup 
 $\omega_0=2.33~\mathrm{eV}$ (green laser), $\theta_0\approx\pi$. The kinematic parameter is then given by $x=8.9$. Assuming accuracy of the Compton edge measurement to be~\cite{Behnke:2013lya} $\Delta\omega/\omega_{max}\lesssim 10^{-3}$, one can expect to be able to test the values $|\kappa| \sim 2\times 10^{-13}$ and, hence, the ratio $|\Delta m_{e}/m_e| \sim 3\times 10^{-9}$ with the LS potential. Similar (slightly improved) sensitivity can be achieved at the planned upgrade of ILC to ${\cal E}=500$~GeV ($x=17.8$) and at CLIC~\cite{CLIC} with ${\cal E}=1.5$~TeV ($x=53.5$).

 If no annual deviation from the nominal Compton edge is found, this would predict that the difference between gravitational and inertial masses of an electron (of positron) will be less than 0.1\%. In analogy to Ref.~\cite{Kalaydzhyan}, if the Compton edge for an electron or positron is measured in two experiments at its nominal position (\ref{nom}) within uncertainties $\Delta \omega_1$ and $\Delta \omega_2$, respectively, then
\begin{align}
\left|\frac{\Delta m_e}{m_e}\right| < \frac{\Delta \omega_1 + \Delta \omega_2}{\omega_{max}}\cdot\frac{m_e^2 (1+x)^2}{4{\cal E}^2 |\Delta \Phi|}\,,
\end{align}
where, as before, $\Delta \Phi$ corresponds to the difference in gravitational potentials for the two experiments.
 As one can also see, the Compton scattering for a positron is independent from the gravitational mass of the electron and vice versa.

{\it Conclusions. ---} We demonstrated a high sensitivity of certain accelerator experiments to the possible violation of WEP for ultrarelativistic massive particles (electrons and positrons). Even though our limits (\ref{numlimits}) on the difference between the gravitational and inertial mass of an electron (positron) are, perhaps, weaker than the ones which can be, probably, obtained from the astrophysical observations~\cite{Mattingly:2005re, Liberati:2013xla,  Tasson:2014dfa, Bluhm:2005uj}, they do not rely on a particular astrophysical model and can be repeated in a well controlled experimental setup. In addition, the limits (\ref{relmlimit}) exploit the long duration of typical accelerator experiments, making it possible to produce results independent of the absolute values of the potentials.

The bounds (\ref{numlimits}, \ref{relmlimit}) can be significantly improved by considering synchrotron losses at LEP~\cite{Altschul:2009xh}\footnote{The modified Lorentz $\gamma$-factor for ultrarelativistic particles following from Eq.~(\ref{energy}), $\gamma_m \simeq \gamma(1+\kappa \gamma^2)$, will affect the synchrotron radiation power, $P\propto \gamma_m^4$, at the circular accelerators.} ($|\kappa|<5\times 10^{-15}$ for electrons and positrons), 1~TeV photons at LHC~\cite{ATLAS:2013ema}, 500~GeV electrons and positrons from ILC~\cite{Behnke:2013lya}, 1.5~TeV electrons and positrons from CLIC~\cite{CLIC}, and 30~TeV photons at HESS~\cite{Klinkhamer:2008ky} ($\kappa < 9\times 10^{-16}$).
This, however, may require a more elaborate analysis, involving additional parameters of the Lorentz-violating Standard Model Extension (SME)~\cite{AmelinoCamelia:2008qg} once the limit on $\kappa$ approaches $10^{-13}$, which is the upper boundary on the next dominating SME parameter~\cite{Kostelecky:2008ts}. For the nonrelativistic antimatter, one can use complementary bounds on the SME parameters coming from bound kinetic energies of the nuclei~\cite{Hohensee:2013hra} and direct spectroscopy~\cite{Kostelecky:2015nma}.

Finally, we proposed laser Compton scattering experiments at the future ILC and CLIC accelerators with estimated sensitivity $|\kappa| \sim 10^{-13}$ improving our limits (\ref{numlimits}, \ref{relmlimit}) by two orders of magnitude.



{\it Acknowledgements. ---} This work was supported in part by the U.S. Department of Energy under Contracts No. DE-FG-88ER40388 and DE-FG0201ER41195. I would like to thank
Dmitry Duev for the astronomical references.




\end{document}